\documentstyle [12pt] {article}
\title{(ANTI)PETER PRINCIPLE - DISCRETE (INVERSE) LOGISTIC EQUATION
WITH IMPRECISELY ESTIMATED AND STIMULATED CARRYING CAPACITY}

\author{Vladan Pankovi\'c$^{\ast}$, Miodrag Krmar$^{\ast}$,Rade Glavatovi\'c$^{\sharp}$\\
$^\ast$Department of Physics, Faculty of Sciences, 21000 Novi
Sad,\\
Trg Dositeja Obradovi\'ca 4. , Serbia, vdpan@neobee.net \\
$^\sharp$ Military-Medical Academy, 11000 Belgrade, Crnotravska
17, Serbia  \\}

\date {}
\begin{document}
\maketitle

\vspace {0.5cm}

\begin {abstract}
In this work we consider the Peter principle and anti-Peter
principle as the discrete logistic and discrete inverse logistic
equation. Especially we discuss imprecisely estimated (by
hierarchical control mechanism) carrying capacity, i.e. boundary
(in)competence level of a hierarchy member. It implies that Peter
principle holds two sub-principles. In the first one objective
boundary competence level is increased for estimation error. In
the second one objective boundary competence level is decreased
for estimation error. Similarly, anti-Peter principle holds two
sub-principles too. All this implies that paradoxical situations
that follow from Peter and anti-Peter principle can be simply
removed by decrease of the error of hierarchical (social) control.
Also we discuss cases by Peter principle when error of the
boundary competence level by estimation grows up. (Then, in fact,
there is no estimation error but stimulation of the boundary level
by control mechanism.) By first Peter sub-principle it implies
anarchy in the social structure or, correspondingly, cancer in the
biology and medicine, schizophrenia in the psychology and
inflation in the economy. By second Peter sub-principle it implies
a totalitary social structure (dictature or caste regime) or
multiplex sclerosis and other autoimmune diseases in biology and
medicine, servile mentality or low value complex in psychology and
depression by hyperactive political influences in economy.
Finally, monotonus changes of the stimulated part of boundary
level cause corresponding phase transitions discussed on the
example of the introspection in the psychology.
\end {abstract}

\vspace{0.3cm}

As it is well-known remarkable Peter principle [1] states,
seemingly paradoxically, that in the hierarchical social structure
any competent member tends to rise to his level of incompetence
beyond its boundary competence level. We observed that form of the
time evolution of member competence level obtained by different
numerical models [2] is very similar to sigmoid form of the
solution of well-known logistic (Verhulst or Maltusian) equation
in the population dynamics. For this reason, in our previous work
[3], we suggested that Peter principle, in the first, rough,
approximation, corresponds to logistic equation with time
increasing population. In this case member competence level
corresponds approximately to population, while boundary competence
level corresponds approximately to carrying capacity. It implies
that here incompetence level is approximately equivalent to
boundary competence level which points out on the limited accuracy
of given approximation. We supposed, without the proof, that a
strict differentiation between incompetence level and boundary
competence level can be done in a better approximation
corresponding by change of the logistic by discrete logistic
equation. Simultaneously we suggest anti-Peter principle that
states that in the hierarchical structure any incompetent member
tends to rise to his level of competence beyond his boundary
incompetence level. It, in a rough approximation that does not
differ boundary incompetence level and competence level, can be
described by inverse logistic equation. It is supposed too,
without the proof, that a strict differentiation between
competence level and boundary incompetence level can be done in a
better approximation corresponding by change of the inverse
logistic by discrete inverse logistic equation.

In this work we shall generalize mentioned our previous results on
the Peter principle and anti-Peter principle. Firstly, we shall
explicitly present discrete logistic and discrete anti-logistic
equation and discuss their solutions. Secondly, it will be shown
that a strict differentiation between incompetence level and
boundary competence level in the discrete logistic equation as
well as a strict differentiation between competence level and
boundary incompetence level needs an additional argument. This
argument represents impreciseness (existence of a non-zero
absolute error) of the estimation (obtained by hierarchical
(social) control mechanism) of the carrying capacity, i.e.
boundary competence and boundary incompetence level of the member.
It implies that Peter principle holds, in fact, two sub-principle.
First one corresponds to situation where objective (accurate)
competent member competence level is increased for estimation
absolute error. In this case competent member can occupy level
over its objective boundary competence level, i.e. given competent
member becomes satisfied but incompetent, in full agreement with
original formulation of the Peter principle. Second one
corresponds to situation where objective (accurate) competent
member boundary competence level is decreased for estimation
absolute error. In this case competent member can occupy only
level under its objective boundary competence level, so that given
member stands competent but unsatisfied. Similarly anti-Peter
principle holds, in fact, two sub-principle too. All this implies
that paradoxical situations that follow from Peter and anti-Peter
principle can be simply removed by increase of the accuracy
(decrease of the absolute error) of the hierarchical (social)
control mechanism at any competence or incompetence level. Also we
shall discuss cases by Peter principle when error of the
estimation grows up. (Then, in fact, there is no estimation error
but stimulation of the boundary level by control mechanism.) By
first Peter sub-principle it implies anarchy in the social
structure or, correspondingly, cancer in the biology and medicine,
schizophrenia in the psychology and inflation in the economy. By
second Peter sub-principle it implies a cast splitting in the
social structure or, correspondingly, sclerosis and dementia in
the biology and medicine, servile mentality in the psychology and
depression in the economy. Finally, monotonous changes of the
stimulated part of boundary level cause corresponding phase
transitions discussed on the example of the introspection in the
psychology.

Thus, as it is well-known logistic (Verhulst or Maltusian)
equation in the population dynamics has form
\begin {equation}
  \frac {dx}{dt} = {\it a}x(1-\frac {x}{r})
  \hspace{1cm} {\rm for} \hspace{0.5 cm} a,r > 0 \hspace{0.5 cm}{\rm and}
  \hspace{0.5 cm} x \leq r
\end {equation}

where $t$ represents the time moment, $x$ - (human or some other
species) population, ${\it a}$  - growth parameter and $r$
carrying  capacity. Simple solution of this equation, representing
a sigmoid function, is
\begin {equation}
 x = x_{0} r \exp[{\it a}t]\frac {1}{r - x_{0} + x_{0}\exp[{\it a}t]}
\end {equation}
where $x_{0}$ represents the initial population smaller than $r$.
Obviously, when $t$ tends toward infinity $x$ tends toward r and
$\frac {dx}{dt}$ toward zero. Given logistic dynamics describes
population growth limited by negative species self-interaction.

It is well known too that there is inverse (anti) logistic
equation (logistic equation for population decrease) corresponding
to (1)
\begin {equation}
    \frac {dx}{dt}= -{\it a}x(\frac {x}{r} - 1)
\hspace{1cm} {\rm for} \hspace{0.5 cm} a,r > 0 \hspace{0.5 cm}{\rm
and} \hspace{0.5 cm} x \geq r
\end {equation}
where $-{\it a}$ represents the decrease parameter. It holds
simple solution
\begin {equation}
   x = x_{0} r \frac {1}{x_{0} - (x_{0} - r) \exp[{\it a}t]}
\end {equation}
where $x_{0}$ represents the initial population smaller than $r$.
Obviously, when $t$ tends toward infinity $x$ tends toward $r$ and
$\frac {dx}{dt}$ toward zero. Given inverse logistic dynamics
describes population decrease limited by positive species
self-interaction.

Finally, it is well-known that both, logistic and inverse
logistic, equations have significant application not only in
population dynamics, i.e. in the biology and demography, but also
in the chemistry, mathematical psychology, economics and
sociology.

It can be added that application of the logistic and inverse
logistic equation needs implicitly that population is sufficiently
large, i.e. effectively continuous. Then change of the population
for few members can be considered as the infinitesimal, i.e.
differential.

But, population can be not so large, i.e. it cannot be considered
as the continuous but as a discrete variable. Then logistic and
inverse logistic equation can be presented in the following
discrete, i.e. finite difference forms with population change as a
finite difference
\begin {equation}
  \frac {\Delta x_{n}}{\Delta t} = {\it a}x_{n-1}(1-\frac {x_{n-1}}{r})
\hspace{0.5cm} {\rm for} \hspace{0.3 cm} a,r > 0 ; \hspace{0.3 cm}
x_{n}= x_{n-1} + \Delta x_{n}\leq r \hspace{0.3cm}{\rm and}
\hspace{0.3 cm} n=1,2,...
\end {equation}
\begin {equation}
  \frac {\Delta x_{n}}{\Delta t}= -{\it a}x_{n-1} (\frac {x_{n-1}}{r}-1)
\hspace{0.5cm} {\rm for} \hspace{0.3 cm} a,r > 0 ; \hspace{0.3 cm}
x_{n}= x_{n-1} + \Delta x_{n}\geq r \hspace{0.3cm}{\rm and}
\hspace{0.3 cm} n=1,2,...
\end {equation}
Solutions of (5), (6), as it is not hard to see, can be presented
in the following inductive forms
\begin {equation}
  x_{n} =  x_{n-1} + ({\it a}\Delta t)x_{n-1} ( 1 - \frac {x_{n-1}}{r}) \geq x_{n-1}
\hspace{0.5cm} {\rm for} \hspace{0.3 cm} a,r > 0 ; \hspace{0.3
cm}x_{n-1}\leq r \hspace{0.3cm}{\rm and} \hspace{0.3 cm} n=1,2,...
\end {equation}
\begin {equation}
  x_{n} = x_{n-1} - ({\it a}\Delta t)x_{n-1} (\frac {x_{n-1}}{r}- 1) \leq x_{n-1}
\hspace{0.5cm} {\rm for} \hspace{0.3 cm} a,r > 0 ; \hspace{0.3
cm}x_{n-1}\geq r \hspace{0.3cm}{\rm and} \hspace{0.3 cm} n=1,2,...
\end {equation}
Here is very important to be pointed out that, for fixed $r$,
limits $x_{n}\leq r$ in (7) and $x_{n} \geq r$ in (8) forbid
appearance of the chaos in given cases.

For $\frac {x_{n-1}}{r}\ll 1$, i.e. for low population levels, (7)
turns out in
\begin {equation}
  x_{n} =  (1 +  ({\it a}\Delta t)) x_{n-1}
\hspace{0.5cm} {\rm for} \hspace{0.3 cm} a,r > 0 ; \hspace{0.3
cm}x_{n-1}\ll r \hspace{0.3cm}{\rm and} \hspace{0.3 cm} n=2,3...
\end {equation}
corresponding to an increasing geometrical progression, i.e.,
intuitively speaking, a pyramidal or hierarchical structure.

But, for $\frac {x_{n-1}}{r}\leq 1$, i.e. for high population
levels, (7) turns out in
\begin {equation}
  x_{n} \geq  x_{n-1}
\hspace{0.5cm} {\rm for} \hspace{0.3 cm} a,r > 0 ; \hspace{0.3
cm}x_{n-1}\geq r \hspace{0.3cm}{\rm and} \hspace{0.3 cm} n \gg 1
\end {equation}
corresponding to a series with almost equivalent values of the
members, i.e., intuitively speaking, to a plateau or saturation
structure.

For $\frac {x_{n-1}}{r}\gg 1$, i.e. for high population levels,
(8) turns out in
\begin {equation}
  x_{n} =  (1 -  ({\it a}\Delta t)) x_{n-1}
\hspace{0.5cm} {\rm for} \hspace{0.3 cm} a,r > 0 ; \hspace{0.3
cm}x_{n-1}\ll r \hspace{0.3cm}{\rm and} \hspace{0.3 cm} n= 2,
3,...
\end {equation}
corresponding to a decreasing geometrical progression, i.e.,
intuitively speaking, an inverse pyramidal or hierarchical
structure.

But, for $\frac {x_{n-1}}{r} \geq 1$, i.e. for small population
levels, (7) turns out in
\begin {equation}
  x_{n} \leq  x_{n-1}
\hspace{0.5cm} {\rm for} \hspace{0.3 cm} a,r > 0 ; \hspace{0.3
cm}x_{n-1}\geq r \hspace{0.3cm}{\rm and} \hspace{0.3 cm} n \gg 1
\end {equation}
corresponding to a series with almost equivalent values of the
members, i.e., intuitively speaking, to a plateau or saturation
structure.

It is not hard to see that discussed characteristics of the
solutions of discrete logistic and inverse logistic equations
excellently correspond to characteristics of the discrete
hierarchical social structure on which Peter or anti-Peter
principle appears.

More precisely, it can be supposed that $x_{n}$ in (5) or (7)
represents the discrete competition level of a member of the
hierarchical social structure.

Suppose now that $r$ representing effective value of the boundary
competence level in (5), (7) can be expressed in the following way
\begin {equation}
  r = r_{0}\pm \delta r
\hspace{0.5cm} {\rm for} \hspace{0.5 cm} r_{0} > \delta r
\end {equation}
Here $r_{0}$ represents the objective (absolute) value of the
boundary competence independent of the estimation of the
hierarchical (social) control mechanism, while $\delta r$
represents the absolute error or inaccuracy of the boundary
competence level estimation done by hierarchical (social) control
mechanism.

For
\begin {equation}
  r = r_{0} + \delta r
\end {equation}
objective boundary competence level is increased for estimation
absolute error. In this case member can occupy level over his
objective boundary competence level (when given member becomes
satisfied but incompetent) in full agreement with original
formulation of the Peter principle. It will be called the first
Peter sub-principle.

For
\begin {equation}
  r = r_{0} - \delta r
\end {equation}
objective boundary competence level is decreased for estimation
absolute error. In this case member can occupy level under his
objective boundary competence level (when given member stands
competent but unsatisfied). It represents a new situation in
respect to original formulation of the Peter principle. It will be
called the second Peter sub-principle.

Analogously, i.e. symmetrically, suppose that r representing
effective value of the boundary incompetence level in (6), (8) can
be expressed in the following way
\begin {equation}
  r = r_{0} \pm \delta r
\hspace{0.5cm} {\rm for} \hspace{0.5 cm} r_{0} > \delta r
\end {equation}
Here $r_{0}$ represents the objective (absolute) value of the
boundary incompetence independent of the estimation of the
hierarchical (social) control mechanism, while $\delta r$
represents the absolute error or inaccuracy of the boundary
incompetence level estimation done by hierarchical (social)
control mechanism.

For
\begin {equation}
  r = r_{0} + \delta r
\end {equation}
objective boundary incompetence level is increased for estimation
absolute error. In this case member can occupy level over his
objective boundary incompetence level (when given member stands
incompetent and satisfied. It represents a new situation in
respect to original formulation of the anti-Peter principle.

For
\begin {equation}
  r = r_{0} - \delta r
\end {equation}
objective boundary incompetence level is decreased for estimation
absolute error. In this case member can occupy level under his
objective boundary incompetence level (when given member becomes
competent but unsatisfied). It is in full agreement with original
formulation of the anti-Peter principle.

In this way it is shown that Peter and anti-Peter principle can be
consistently formulated by discrete logistic and discrete inverse
(anti) logistic equation with effective (objective changed by
estimation error) boundary level of the competence or
incompetence. It implies that paradoxical situations that follow
from Peter and anti-Peter principle can be simply removed by
increase of the accuracy (decrease of the absolute error) of the
hierarchical (social) control mechanism.

Now, we shall discuss what happened when absolute error of the
estimation dr grows up during time in the discrete logistic
equation, i.e. by Peter principle (7).

Firstly we shall consider first Peter sub-principle, described by
(7) and (14). Suppose that given absolute error grows up
monotonously during time. For reason of the mathematical
simplicity it can be considered that time interval of the
competence level evolution can be divided in many sub-intervals.
In any of given time sub-intervals discrete logistic equation with
constant boundary competence level can be applied, but in the
different time sub-intervals corresponding boundary competence
levels will be different too. Thus, after some time interval
absolute error and, according to (14), whole effective boundary
competence level, can be many times larger than practically any
competence level. Obtained discrete logistic equation becomes
formally equivalent to (9) and describes an unlimitedly increasing
geometrical progression. Meanwhile, there is a principal
difference between obtained equation and (9). Namely, in obtained
equation absolute error is dominant in respect to objective
boundary competence level, while in (9) objective boundary
competence level is dominant in respect to absolute error. In
other word (9) describes a maximally functional and maximally
self-controlable hierarchical structure. This structure holds
correctly defined and strictly different discrete competence
levels necessary for successful functioning. Also (9) describes
optimal motion of any competent member up hierarchical (social)
structure, according to his competence. Vice versa, obtained
equation describes practically a totally anarchical (chaotical)
social structure without any self-control. This singular social
structure represents a chaotic mixture of all competence levels.
It means that any member can occupy any position without any
criterion. For this reason functioning of given singular structure
is completely destroyed.

It is not hard to see that given unlimitedly expanding anarchy
(chaos) in the social structure conceptually corresponds to cancer
in the biology and medicine, schizophrenia in the psychology and
inflation in the economy. In any of given domains, as it is
well-known, decline of corresponding self-control and integrative
mechanism implies catastrophic consequences.

Now, we shall consider second Peter sub-principle, described by
(7) and (15). Like in the previous case, it will be supposed that
absolute error, initially smaller than objective boundary
competence level, grows up monotonously during time. Then, during
time, effective boundary competence level, according to (15),
becomes smaller and smaller, tending toward zero. In the time
moment when effective boundary competence level becomes smaller
than practically any competence level (7) turns out in the
equation formally equivalent to (8). (Simply speaking, Peter
principle becomes formally inverted in the anti-Peter principle.)
Meanwhile, there is a principal difference between obtained
equation and (8). Namely, obtained equation is consequence of the
large absolute error and (15), while in (8) objective boundary
competence level is dominant in respect to absolute error. In
other words, (8) describes a maximally functional and maximally
self-controlable hierarchical structure, with correctly defined
and strictly different discrete competence levels. Also (8)
describes optimal motion of any incompetent member down
hierarchical (social) structure, according to his incompetence.
Vice versa, obtained equation describes practically a totalitary
social structure (dictature or caste regime). This structure holds
extremely separated competence levels (like in a caste system). It
means that any competent member is far away under his objective
competence level for reason of the hyperactive and destructive
hierarchical self-control (pressure).

It is not hard to see that given totalitary social structure
(dictature or caste regime) conceptually corresponds to multiplex
sclerosis and other autoimmune diseases in the biology and
medicine, servile mentality or low value complex in the psychology
and depression by hyperactive political influences in the economy.

Finally consider complete situation where control mechanism
influence, i.e. stimulated boundary competence level changes
monotonously during time. It corresponds to a phase transitions
from one into other limit, from dictature to anarchy or vice
versa. As it is not hard to see given transition can be
demonstrated on the introspection phenomenon in the psychology.
Metaphorically, introspection represents the sinking of the
consciousness in the non-consciousness. More precisely,
introspection (self-observation) represents, estimation of the
irrational (that increases) by rational (that decreases) part of a
human mind, or, increasing low integrated by decreasing high
integrated functions of the brain neural system.

In conclusion we can shortly repeat and point out the following.
Logistic or inverse logistic (in differential or finite difference
form) equation represents one of the most general equation that
describes simply (negative quadratically) self-controled system in
simple (linear) interaction with (positively or negatively)
influenced environment. For this reason given equation holds a
great domain of the application: in the biology, medicine,
psychology, sociology and economy. Peter or anti-Peter principle
(with two sub-principles) points out that logistic or inverse
logistic equation holds implicitly a more general form within
which boundary carrying capacity can be treated as a variable, but
not as a constant parameter. This variation of the boundary
carrying capacity represents result of the influence (positive or
negative stimulation) realized by control mechanism. In this sense
we can speak about objective boundary carrying capacity
representing a constant independent of the control mechanism
influence and stimulated part of the boundary carrying capacity
representing a variable dependent of the control mechanism
exclusively. When stimulated part of the boundary carrying
capacity becomes smaller and smaller (when it can be simply
treated as the control mechanism estimation error) paradoxical
situations characteristic for Peter and anti-Peter principle
disappear. In the limit of minimal value of stimulated part of
boundary carrying capacity we obtain a maximally functional
hierarchical structure (with optimally placed any member). In the
opposite case, when stimulated part of the boundary carrying
capacity becomes larger and larger, paradoxical situations
characteristic for Peter and anti-Peter principle becomes more and
more expressed. Then, for positive stimulated part of the boundary
carrying capacity, an anarchical situation (total destruction of
the functional hierarchical structure) appears. (It corresponds to
anarchy in the social structure, cancer in the biology and
medicine, schizophrenia in the psychology and inflation in the
economy.) Or, for negative stimulated part of the boundary
carrying capacity, a sclerotic situation (with a total
de-acceleration of the functioning of the hierarchical structure)
appears. (It corresponds to a totalitary social structure
(totalitary or caste regime), multiplex sclerosis and other
autoimmune diseases in the biology and medicine, servile mentality
or low value complex in the psychology and depression by
hyperactive political influences in the economy.) Finally,
monotonous changes of the stimulated part of boundary carrying
capacity cause corresponding phase transitions discussed on the
example of the introspection in psychology.

\vspace{2.5cm}

{\Large \bf References}

\begin {itemize}

\item [[1]] L. J. Peter, R. Hul, {\it The Peter Principle: why things always go wrong} (William Morrow and Company, New York, 1969)
\item [[2]] A. Pulchino, A. Rapisarda, C. Garofalo, {\it The Peter Principle Revisited: A Computational Study},
soc-ph/0907.0455 and references therein
\item [[3]] V. Pankovic, {\it Peter and Anti-Peter Principle as the Discrete Logistic
Equation},\\
gen-phys/0907.2509 v1

\end {itemize}

\end {document}